# Calibration and Status of the 3D Imaging Calorimeter of DAMPE for Cosmic Ray Physics on Orbit

Libo Wu, Sicheng Wen, Chengming Liu, Haoting Dai, Yifeng Wei, Zhiyong Zhang, Xiaolian Wang, Zizong Xu, Changqing Feng, Shubin Liu, Qi An, Yunlong Zhang, Guangshun Huang, Yuanpeng Wang, Chuan Yue, JingJing Zang, Jianhua Guo, Jian Wu, Jin Chang

*Abstract*—The DArk Matter Particle Explorer (DAMPE) developed in China was designed to search for evidence of dark matter particles by observing primary cosmic rays and gamma rays in the energy range from 5 GeV to 10 TeV. Since its launch in December 2015, a large quantity of data has been recorded. With the data set acquired during more than a year of operation in space, a precise time-dependent calibration for the energy measured by the BGO ECAL has been developed. In this report, the instrumentation and development of the BGO Electromagnetic Calorimeter (BGO ECAL) are briefly described. The calibration on orbit, including that of the pedestal, attenuation length, minimum ionizing particle peak, and dynode ratio, is discussed, and additional details about the calibration methods and performance in space are presented.

*Index Terms*—DAMPE, BGO, Calorimeter, Cosmic rays.

## I. INTRODUCTION

THERE are some space-borne experiments (AMS [1], Fermi-LAT [2], PAMELA [3], CALET [4]) to search for evidence of dark matter by measuring the spectra of photons, electrons and positrons. Exciting results from the successful operation of these experiments have been published [5-8]. The DArk Matter Particle Explorer (DAMPE) is an orbital experiment supported by the Strategic Priority Science and Technology Projects in Space Science of the Chinese Academy of Sciences. It was successfully launched into a sun-synchronous orbit at the altitude of 500 km on 17[th] December 2015 from Jiuquan launch center.

The BGO Electromagnetic Calorimeter (BGO ECAL) is one of the key sub-detectors of DAMPE [9]. It provides good energy resolution (≤ 1.5% at 800 GeV for e/γ [9]) and high electron/hadron separation (>$10^5$). During more than a year, the BGO ECAL has been in operation as expected, providing the trigger to the DAMPE detector and measuring cosmic ray energy. To exploit the full potential of the BGO ECAL in space environment, a dynamic BGO ECAL calibration method was developed.

## II. THE DAMPE DETECTOR

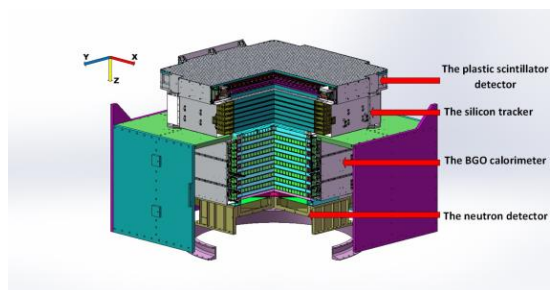

Fig. 1. Side view of the DAMPE detector

The layout of DAMPE detector is shown in Fig. 1. DAMPE is equipped with four different sub-detectors.

The Plastic Scintillator strip Detector (PSD) [9] has a 825×825 $mm^2$ active area of organic plastic scintillator, and it is arranged in a double layer configuration and has in total 82 detector modules. The PSD is designed to measure the charge (Z) of incident high-energy particles up to Z = 26, and it is also used to identify electrons and gamma rays.

The Silicon-Tungsten tracKer (STK) [9, 11] consists of six planes (with each plane having two orthogonal layers) of silicon microstrip detectors and three tungsten plates with 1.0 mm thickness. The three tungsten plates are inserted in front of tracking planes 2, 3, and 4 for photon conversion. The STK is devoted to the precise reconstruction of the particle track and e/γ identification. The spatial resolution is better than 80 μm within 60° incidence [9].

The BGO calorimeter is used to measure the energy deposition of e/γ with a good resolution of 1.5% for 800 GeV incident particles and to reconstruct the shower profile, which is fundamental to distinguishing electromagnetic showers from hadronic showers. Moreover, the trigger of the whole DAMPE system is based on the signals from the BGO.

The Neutron Detector (NUD) [9, 12] at the bottom of DAMPE aims to distinguish the types of high-energy showers

The DAMPE mission is funded by the Strategic Priority Science and Technology Projects in Space Science of the Chinese Academy of Sciences. This work is supported by the National Key Research and Development Program of China (2016YFA0400200 and 2016YFA0400202), the Project supported by the Joint Funds of the National Natural Science Foundation of China (No. U1738208, U1738139 and U1738135) and the National Natural Science Foundation of China (No. 11673021 and No. 11705197).

L. B. Wu, Y. L. Zhang, Y. F. Wei, Z. Y. Zhang, C. M. Liu, H. T. Dai, X. L. Wang, Z. Z. Xu, G. S. Huang, C. Q. Feng, S. B. Liu and Q. An are with the State Key Laboratory of Particle Detection and Electronics & Department of Modern Physics, University of Science and Technology of China, Hefei, Anhui 230026, China (e-mail: ylzhang@ustc.edu.cn).

S. C. Wen, Y. P. Wang, C. Yue, J. J. Zang, J. H. Guo, J. Wu, J. Chang are with the Purple Mountain Observatory, Chinese Academy of Sciences, Nanjing 210008, China.

to improve e/p identification capacity. It consists of four boron-loaded plastics, each with a set of photomultiplier tubes (PMTs), and an auxiliary circuit.

## III. THE BGO ECAL

The role of the calorimeter is threefold:
- Measuring the energy deposition of electron, positron and photon over a wide energy range from 5 GeV to 10 TeV.
- Imaging the three-dimensional profile for electron/hadron shower.
- Providing the level 0 trigger for the DAMPE data acquisition system [13-15].

### A. Three-dimensional imaging calorimeter

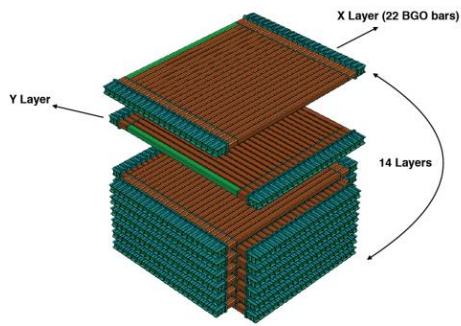

Fig. 2. Enlarged view of the BGO ECAL

DAMPE's BGO ECAL is a total absorption calorimeter consisting of 308 BGO crystal [16] bars that allow for precise three-dimensional imaging of the shower shape. The ECAL has 14 layers, giving it about 32 radiation lengths and 1.6 nuclear interaction lengths. Each layer contains 22 BGO crystal bars with dimensions of $25 \times 25 \times 600$ mm$^3$. The crystal bars are located in a honey comb-like carbon fiber cage. The BGO crystal bars of neighboring layers are alternated in an orthogonal way to measure the deposited energy and image the shower profile developed in the BGO ECAL in both views ($x$ – $y$ and $y$ – $z$) (Fig. 2). Furthermore, the readout channels of the first and the last 4 layers of the BGO ECAL are used to provide signals for the following DAMPE trigger logic: MIP, high energy, low energy and unbiased. The key parameters of the BGO ECAL are listed in Table I.

TABLE I
BGO SPECIFICATIONS

| Parameter | Value |
|---|---|
| Active area | $60 \times 60$ cm$^2$ (on-axis) |
| Radiation lengths | 32 X$_0$ |
| Nuclear interaction length | ~ 1.6 λ$_I$ |
| Sampling | > 90% |
| Longitudinal segmentation | 14 layers |
| Energy loss (for MIPs) | 9.2 MeV/cm |

### B. Light collection and readout system

Two sides of a crystal bar are coupled to R5610A-01 Hamamatsu PMTs (named S0 and S1, respectively), which collect light of BGO. To measure the energy from 5 GeV to 10 TeV, a multi-dynodes readout circuit of PMT was designed [15]; the signals are read out from the different sensitive dynodes 2, 5, and 8 (Dy2, Dy5, Dy8) corresponding to low, middle, and high gain. (Fig. 3). The front-end electronics (FEE) is outfitted with a VA160 or VATA160 chip [17] whose dynamic range is from -3 pC to +13 pC developed by IDE AS and is composed of a charge-sensitive preamplifier, a CR-RC shaping amplifier, and a sample-hold circuit.

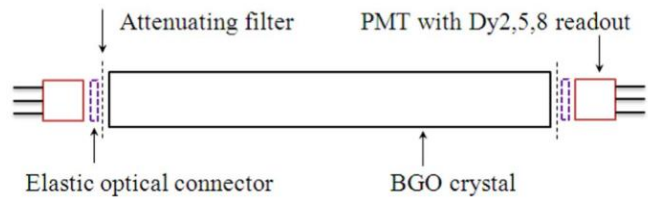

Fig. 3. Minimum detection unit of the BGO ECAL

### C. Performance of the hardware in space

To assess the BGO ECAL operational capability in space, all the components underwent several environmental tests [18], including magnetic moment measurement, electromagnetic compatibility, vibration, and thermal vacuum tests. In addition to that, the BGO ECAL has been space qualified, passing through all strict environmental tests. To ensure long-term reliability in space, a certain degree of redundancy has been implemented in the BGO ECAL: Each detecting unit is connected to two PMTs, and each PMT has three readout channels. During more than a year of data taking on orbit, the BGO ECAL worked as expected without any major issue.

## IV. BGO ECAL CALIBRATION

The BGO ECAL detector provides precise measurement of the energy of e/γ. There are several calibration steps, including pedestal, dynode ratio, attenuation and MIPs. Pedestal set "zero point" for energy scale, MIP calibration can give an absolute energy scale for dynode 8, and dynode ratio can be used to convert the scale of dynode 8 to ones of dynode 5 and 2, attenuation length can be used to correct energy measured which is position dependence due to scintillation light attenuation in BGO bar. In the following sections, all the calibrations are described and the performance of the BGO ECAL in space is also reported.

### A. Pedestal calibration

Pedestals are offset voltages for each of the readout unit that set the "zero point" for the energy scale. They are calibrated at latitude of 20°N from the periodic triggers issued at 100 Hz by the DAMPE trigger system. There are 2016 electronic channels used in the calorimeter, 1848 of them are connected to the dynodes of the PMTs for signal readout, while the others are floated for monitoring the electronic chains. The peak of pedestal spectrum reflects the reference counts of each ADC channel, and the width of pedestal reflects the noise of the readout channel. Fig. 4 shows the typical pedestal spectra of Dy2, Dy5, and Dy8 fitted by Gaussian functions. Fig. 5 reports the pedestal widths. Noise of the 1848 channels connected to the BGO calorimeter is about 7 ADC channels, corresponding to ~ 5 fC injected into the FEE chips. Channels whose noise is lower than 4 ADC counts are spare ones that were not connected to the PMTs. Fig. 6 shows small variations in pedestal sigma and mean, indicating a stable readout noise in the space. In order to study the change of pedestal at different latitude, a monitoring of the pedestal had been done after the launch. Fig. 7 shows the noise behavior of one channel, and the results show that the mean value has a large drop at the moment when DAMPE passes through South Atlantic Anomaly (SAA) area; furthermore, the pedestal also undergoes a small drop in the outer radiation belt. The data collected at these areas is not use for analysis.

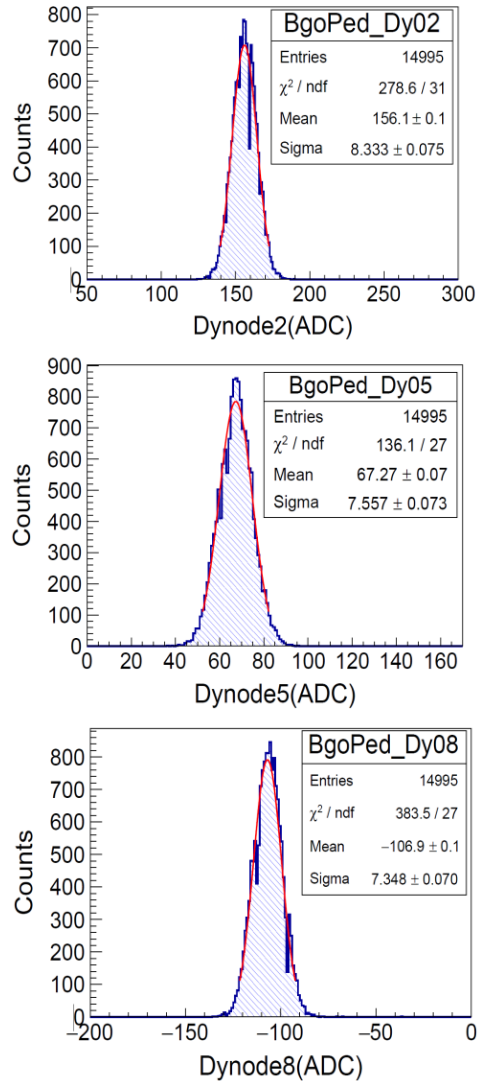

Fig. 4. Typical pedestal spectra of dy2, dy5, and dy8

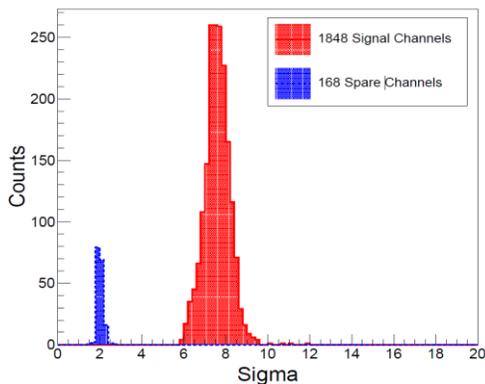

Fig. 5. Distribution of the pedestal sigma values of all channels

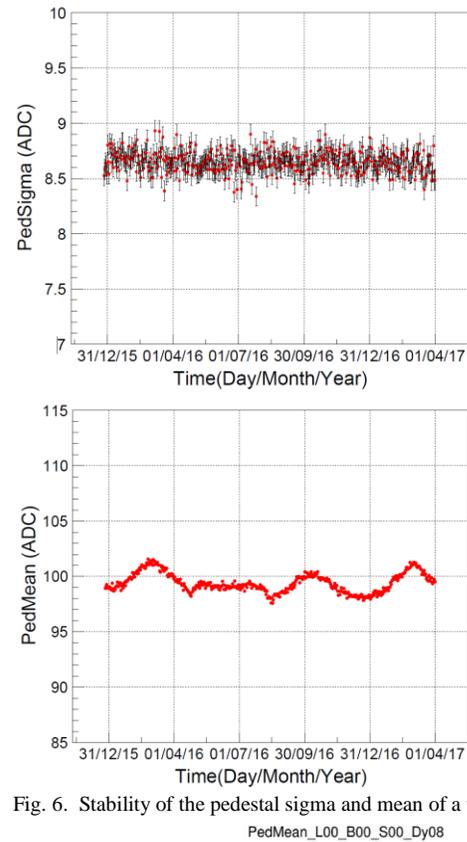

Fig. 6. Stability of the pedestal sigma and mean of a typical channel

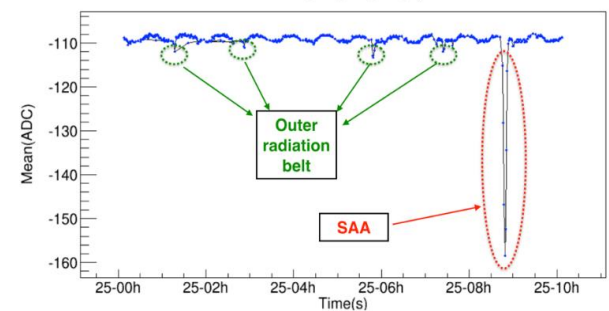

Fig. 7. Short-time stability of the mean pedestal counts of a typical channel

B. *Dynode ratio calibration*

TABLE 2
THE ENERGY RANGE OF A BGO CRYSTAL

|  | Side0 | Side1 |
| --- | --- | --- |
| Dynode8 | 2 MeV ~ 500 MeV | 10 MeV ~ 2500 MeV |
| Dynode5 | 80 MeV ~ 20 GeV | 400 MeV ~ 100 GeV |
| Dynode2 | 3.2 GeV ~ 800 GeV | 16 GeV ~ 4000 GeV |

As mentioned before, to cover a large dynamic range from 5 GeV to 10 TeV in the BGO ECAL, the PMTs are designed in a multi-dynode readout structure. In order to further extend the dynamic range, the light generated in the BGO crystals are read out unequally from the two readout sides by applying attenuation filters with different factors (Fig. 3.). The ranges of different readout dynodes are listed in Table 2. So the most important thing is each readout ADC scale of dynode 5 and dynode 2 should be converted into corresponding ADC scale of dynode 8, and the ratios between the three dynodes are key parameters for energy reconstruction. The shower events in BGO ECAL are used for dynode ratio calibration. The results are shown in Fig. 8. The dynode ratios of Dy5/Dy8 and Dy2/Dy5 are both about 0.02, which means that the energy scale of Dy2 is about 50 times that of Dy5's and 2500 times that of Dy8's. The PMTs of the BGO ECAL are surrounded by permalloy to shield them from Earth's magnetic field. As shown in Fig. 9, the ratio of Dy5/Dy8 does not vary with latitude, confirming the effectiveness of the magnetic shielding. Fig. 10 shows that the ratios of Dy8/Dy5 and Dy5/Dy2 do not depend on the running time for about one year. The results demonstrate that the PMTs and the FEE have been in proper working condition.

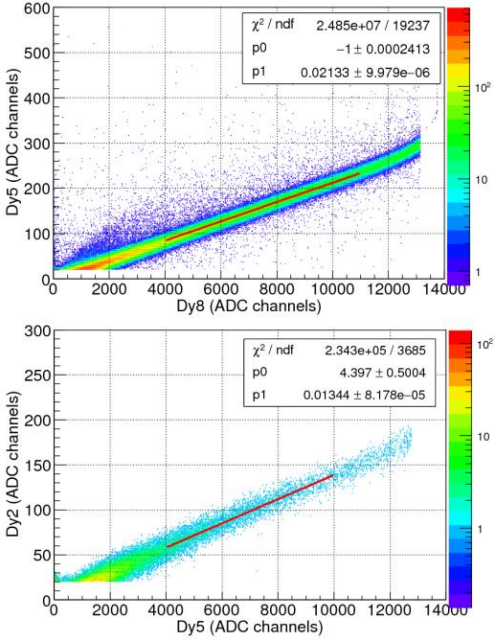

Fig. 8. Dynode ratios of Dy5/Dy8 and Dy2/Dy5, obtained from Side0

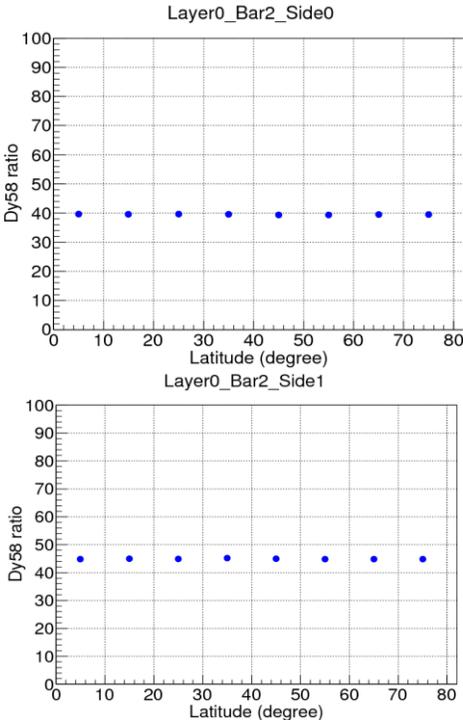

Fig. 9. Dynode ratios, read out from the two sides, as function of the Earth's latitude

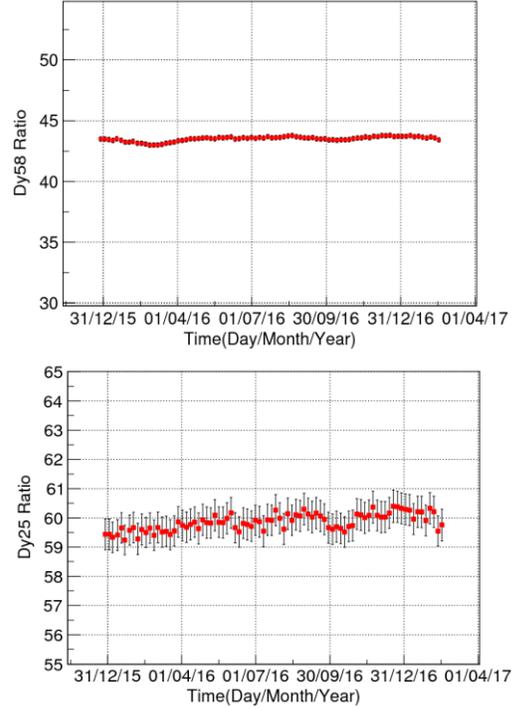

Fig.10. Stability of dynode ratios Dy8/Dy5 (up) and Dy5/Dy2 (down)

*C. Attenuation calibration and spatial correction*

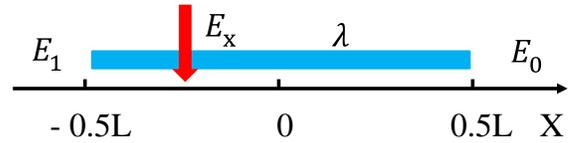

Fig. 11. An illustration of attenuation calibration

Due to the specific 60-cm long BGO crystal, the exponential attenuation of the scintillation light should be taken into account when it transmits along the axis of the BGO crystal bar. The attenuation length is calibrated with shower events. As shown in Fig. 11, the energy deposition of incident particle at the position x is $E_x$; the scintillation light yield produced at the position x is represented as $Ly_x$, which should satisfy the relation of $Ly_x=kE_x$ (k is a constant) when there is no fluorescence saturation in BGO crystal bars. Due to the attenuation of light propagating in BGO bar, the relation between $E_x$ and the energies measured from side0 ($E_0$) and side1 ($E_1$) should be:

$$E_0 = E_x e^{-(\frac{L}{2}-x)/\lambda} \quad (1)$$

and

$$E_1 = E_x e^{-(\frac{L}{2}+x)/\lambda} \quad (2)$$

where $x$ is the hit position of the particle, which varies from

−300 mm (Side0) to 300 mm (Side1) and is determined by STK tracks; $\lambda$ is representative of the attenuation length of the BGO crystal; and $L$ is the length of the crystal (600 mm). From formulas (1) and (2). We can obtain the relationship between $x$ and $\lambda$:

$$\ln\left(\frac{E_0}{E_1}\right) = \frac{2x}{\lambda} \quad (3)$$

We can therefore calibrate $\lambda$ for a given BGO bar with the electromagnetic shower events which have well defined energy deposit and hit positions x located by STK and BGO ECAL itself .The data ADC for both side 0,1 are readout, $E_0$ and $E_1$ are calculated using their own energy scales. After more than a year data taking on orbit, the variation of $\ln\left(\frac{E_0}{E_1}\right)$ as a function of the hit position x is shown in Fig. 12. After fitting by a linear function (the formula 3), the attenuation length of each BGO crystal bar can be obtained.

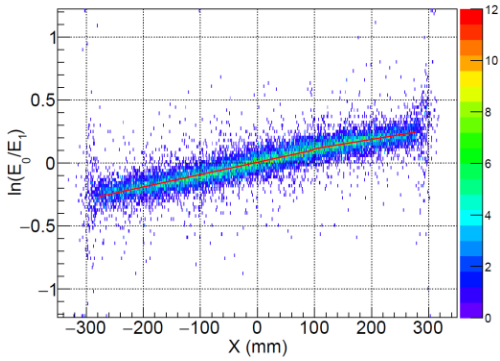

Fig. 12. Scatter plot of $\ln\left(\frac{E_0}{E_1}\right)$ versus $x$ before correction

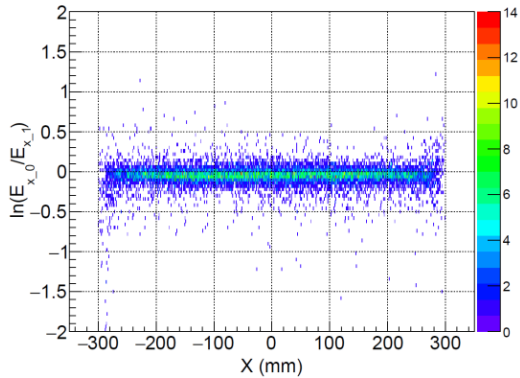

Fig. 13. Scatter plot of $\ln\left(\frac{E_{x\_0}}{E_{x\_1}}\right)$ versus $x$ after correction

After attenuation calibration, the $\lambda$ value of each bar can be used for energy correction of each basic readout unit for every incident event. The correction energy for the two side can be expressed as

$$E_{x\_0} = E_0 e^{(\frac{L}{2}-x)/\lambda} \quad (4)$$

and

$$E_{x\_1} = E_1 e^{(\frac{L}{2}+x)/\lambda} \quad (5)$$

where $E_{x\_0}$ and $E_{x\_1}$ are the correction energy values for the two sides that are independent of the incident position. The resulting variation of $\ln\left(\frac{E_{x\_0}}{E_{x\_1}}\right)$ with respect to the hit position $x$ is plotted in Fig. 13. By comparison with Fig. 12, we see that the curve is flatter after correction.

### D. *MIP calibration*

The response of the BGO ECAL to MIPs is the reference of energy reconstruction, and the precision of the MIP calibration influences the energy reconstruction directly. On the ground, cosmic muons, beam hadrons and muons can be used to calibrate the MIP responses; however, the instrument can only be calibrated by MIP protons or MIP light ions on orbit. Due to the earth's magnetic cut-off, the energy spectra of cosmic proton are well known around 10 GeV between the latitudes of 20°N and 20°S, so the data sample of proton events collected in those attitudes should be suitably chosen for BGO-ECAL calibration. MIP events were selected according to the following logic:

- Crossing cut: The first or second layer and the thirteenth or fourteenth layer must have signals to select the event that crosses the whole BGO ECAL.
- Energy cut: The total energy deposited in the BGO ECAL must be less than 50 MIPs. (The MIPs unit was calibrated in the last time; the first MIPs unit was calibrated on the ground.)
- Hits cut: The number of layers with no signal must be less than 4. Furthermore, the number of BGO crystal bars that have signal must be less than 2 for each layer.
- Track cut: There is only one track in the STK.
- Track fitting cut in the BGO: The chi-square fitting with the track must be less than 2.7 in the BGO ECAL.

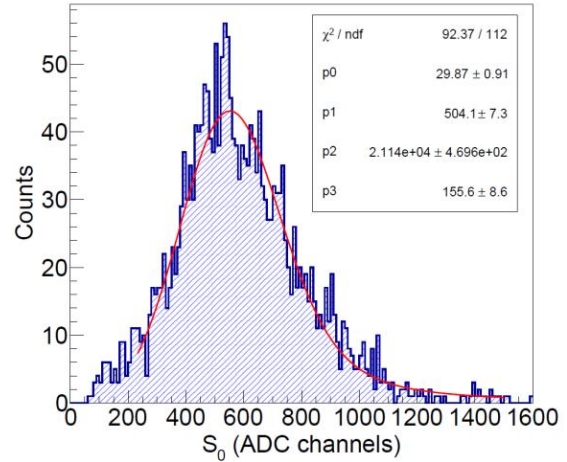

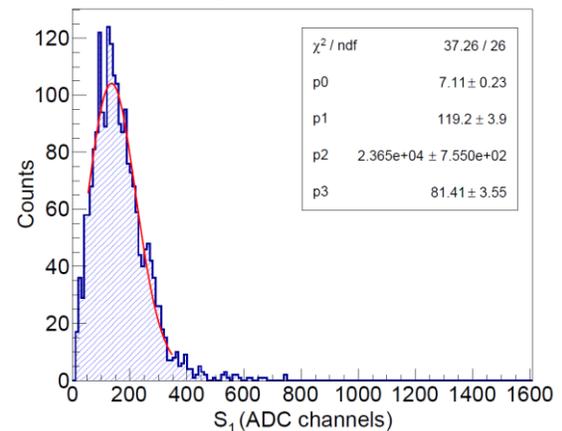

Fig. 14. MIP spectra of Side 0 (up) and Side 1 (down)

Typical MIP spectra, fitted by a Landau convoluted Gaussian function, from the two sides of the crystal are shown in Fig. 14. Due to the attenuation filters with different factors (Fig. 3.), the spectra of side0 and side1 are different.

BGO is a kind of inorganic crystal, and its scintillation light yield is strongly temperature dependent. The reference value of temperature dependence of a BGO crystal is -0.9% / ℃[19], and the value of PMT's anode is -0.4% / ℃[19], which will influence the light output directly. Limited by statistics of MIP events in space, it takes at least 3 days to calibrate the MIP response. The stability performance is shown in Fig. 15. The most probable value (MPV) was used to represent the light output of BGO crystal. Its range is from 560 ADC channels to 620 ADC channels, which is inversely related to the variation of temperature as expected [19]. After temperature correction and attenuation correction, the energy of a MIP event is about 312 MeV, and the stability of energy measurement is better than 1%.

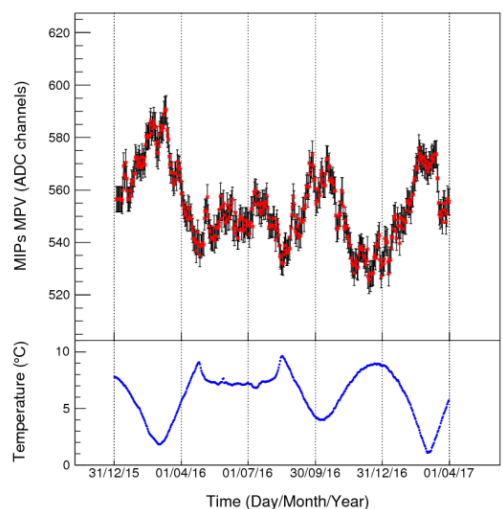

Fig. 15. Stability of the MPV value for MIP proton events

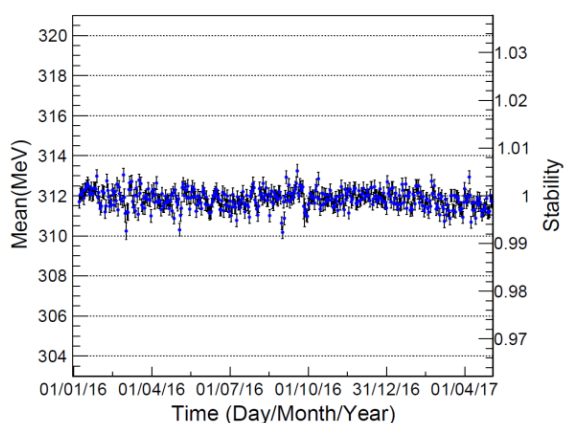

Fig. 16. Stability of the energy for MIP proton events after temperature correction

## V. CONCLUSION

The BGO ECAL plays a major role in the DAMPE experiment for studying cosmic ray physics. Since December 2015, the BGO ECAL has been operating in space, and all detecting units are working properly. The Parameters of BGO ECAL are very stable on orbit for more than a year, including pedestal, MIPs, attenuation length and dynode ratio.